# An *Ab Initio* Study on Energy Gap of Bilayer Graphene Nanoribbons with Armchair Edges

Kai-Tak Lam[a] and Gengchiau Liang[b]

Department of Electrical and Computer Engineering, National University of Singapore

Dependency of energy bandgap ($E_g$) of bilayer armchair graphene nanoribbons ($\text{AGNR}_B$) on their widths, interlayer distance ($D$) and edge doping concentration of boron/nitrogen is investigated using local density approximation and compare to the results of monolayer graphene nanoribbons ($\text{AGNR}_M$). Although $E_g$ of $\text{AGNR}_B$, in general, is smaller than that of $\text{AGNR}_M$, $\text{AGNR}_B$ exhibits two distinct groups, metal and semiconductor, while $\text{AGNR}_M$ displays purely semiconducting behavior. $E_g$ of $\text{AGNR}_B$, moreover, is highly sensitive to $D$, indicating a possible application in tuning $E_g$ by varying $D$. Finally, edge doping of both AGNR systems reduces $E_g$ by 11-17%/4-10% for $\text{AGNR}_M$/$\text{AGNR}_B$, respectively.

[a] lamkt@nus.edu.sg
[b] elelg@nus.edu.sg

Carbon related materials recently have generated much interest due to their unique physical, electronic, and optical properties. Carbon atoms arranged in a linear honeycomb structure in a two dimension (2D) plane form graphene sheets which are the basis for many other carbon nanostructures such as fullerenes, carbon nanotubes and graphene nanoribbons. Recent experiments[1-3] show the possibility to cut or pattern the 2D graphene sheet into nanometer width strips named graphene nanoribbons (GNRs). GNRs have different electrical properties compared to graphene sheets due to quantum confinement as well as the abrupt terminations at the edges. From tight-binding[4-6] and first-principles calcuations[7-10], these GNRs are semiconducting materials with an energy gap ($E_g$) dependent on the ribbon width as well as the atomic configuration of the edges. More specifically, for GNRs with armchair edges, the $E_g$ decreases with increasing width, exhibiting 3 distinct trends in the $E_g$ variations. Application of GNR in field-effect transistor devices has also been studied theoretically[11-15] and experimentally[1-3]. In order to keep the high performance, doped source and drain in field effect transistors are required. GNRs provide the possibility to substitute the edge carbon atoms with either boron or nitrogen which will lead to a *p*-type or *n*-type doped semiconductor, respectively[12,13,16]. Apart from monolayer GNRs, recent experimental[17-19] and theoretical[20] studies are also carried out based on bilayer graphene and they show that bilayer graphene has unique features such as anomalous integer quantum Hall effects[17] and the ability to control the size of $E_g$ by adjusting carrier concentration[18] as well as by an external electric field[19]. These unique properties open up the opportunity to implement bilayer GNRs in the various applications.

Understanding the stable geometry of bilayer GNRs, their electronic structures, and their fundamental physics is the essential steps towards device realization. In this letter, therefore, we investigate the electronic structure of armchair-edge bilayer GNRs (AGNR$_B$) and the dopant effects on bandgap using first-principles calculations based on Density Functional Theory (DFT) within local density approximation, implementing the self-consistent pseudopotential method and the double-ζ (polarized) basis set by the conventional software package, Atomistix ToolKit 2.2[21-23]. We first determine the optimum interlayer distance of AGNR$_B$ based on the total energy calculations as function of the interlayer distance. We found that the optimum interlayer distance of AGNR$_B$ is smaller than that of bilayer graphene due to non-uniform charge distribution in AGNR$_B$. Then, we examine the effect of interlayer distance on $E_g$ of AGNR$_B$, followed by the width dependency of $E_g$ of AGNR$_B$ and contrast it with that of armchair-edge monolayer GNRs (AGNR$_M$). Like AGNR$_M$, AGNR$_B$ also shows three different groups in terms of $E_g$ dependence on width, and in general, AGNR$_B$ is found to have a lower $E_g$ than AGNR$_M$. Especially for $N$=3p+2 family, while it is semiconducting for AGNR$_M$[8], the $E_g$ of AGNR$_B$ is very small and can be considered as metallic at room temperature. Furthermore, we investigate the relationship between $E_g$ and the interlayer distance ($D$) of AGNR$_B$, and $E_g$ is found to be strongly influenced by $D$. Finally, we present the edge doping effect on both AGNR$_M$ and AGNR$_B$, which are found to form *p*-type and *n*-type semiconductor with boron and nitrogen.

Following the nomenclature in Ref. 8 and references within, the width of the AGNR$_M$ is related to the number of dimmer lines ($N$) across the ribbon width. The honeycomb structure, with hydrogen-passivation, of AGNR$_M$ with $N$=7 (or 7-AGNR$_M$) is shown in Fig. 1(a). The carbon-carbon bond and the carbon-hydrogen bond are set at 1.42Å and 1.09Å, respectively and the structure is allowed to relax until the forces between the atoms are less than 0.05eV/Å. The boxed regions in Fig. 1(a) indicated the sites where boron and nitrogen atoms are more energetically favorable to replace a carbon atom[12].

In a bilayer graphene, the two graphene sheets are arranged with the atoms in one layer located on top of the center of a hexagon in the other layer (shown in Fig. 1(b)). This arrangement is commonly found in nature[24] and previous calculation had shown that it is the most energetically favorable configuration[25]. The total energy of an $AGNR_B$ is calculated by stacking 2 $AGNR_M$ in similar fashion in the simulation. The total energy of a structure where the $AGNR_M$ are directly above each other is also calculated and the results show that the former structure has a lower total energy, indicating a more energetically favorable $AGNR_B$. Similar to the nomenclature of $AGNR_M$, a 7-$AGNR_B$ is shown in Fig. 1(c).

In order to determine the optimum interlayer distance ($D$) between the GNR in an $AGNR_B$, the total energy of a 7-$AGNR_B$[26] is calculated as $D$ varies. For comparison, the total energy of a bilayer graphene[26] with varying $D$ has also been calculated and the normalized results shown in Fig. 2(a). The results show that there are two stable energy points for both bilayer graphene and $AGNR_B$ as $D$ varies. For $D$>4.9Å, the total energy become a plateau as $D$ varies, indicating the interaction between the layers is negligible at a distance larger than 4.9Å. For $D$<4.9Å, the interlayer interactions become significant and the total energy of the system quickly reaches a minimum at $D$≈3.3Å for bilayer graphene. It is in a good agreement with previous first-principles calculation[25] which showed that the interlayer distance for bilayer graphene was around 3.4Å. Using 7-$AGNR_B$ as an example, we found that the minimum energy occurs at $D$≈3.1Å. To investigate the variation on interlayer distance between infinite bilayer graphene and $AGNR_B$, the charge density in the plane between the layers for bilayer graphene and 7-$AGNR_B$ are plotted in Fig. 2(b) and (c) respectively. For bilayer graphenes, the charges are concentrated at the points where the carbon atoms of the two layers are aligned, and periodically distributed along the entire plane. Comparatively, the charges are not as strongly localized at the aligned carbon atoms for 7-AGNR$_B$, but there are more charge accumulation at the edges of 7-AGNR$_B$. Furthermore, we found that as AGNR$_B$ width increase, the optimum $D$ increases, as shown in the insert of Fig. 2(a). This suggested that edge-effect plays an important role in charge distributions and in determining the optimum $D$ of an AGNR$_B$.

Next, the energy gap ($E_g$) dependency on the width ($W$) of AGNR$_B$ is investigated, as shown in Fig. 3(a). We first examine the case where $D$=6.5Å (hollow points) where the interlayer interaction is not significant following the calculations above. The $E_g$ of this system shows three families ($N$=3p, 3p+1 3p+2, p is a positive integer) which exactly match the monolayer trends (dot-dashed lines). It indicates that an AGNR$_B$ with large $D$ can be treated as two AGNR$_M$ in terms of electronic structure. On the other hand, at their respective optimum $D$ ($D_{op}$) which is different at different width of AGNR$_B$, the $E_g$ of AGNR$_B$ (solid points in Fig. 3(a)) is smaller by 32-96% as compared to that of AGNR$_M$. More specifically, for the family of $N$=3p+2, the $E_g$ of AGNR$_B$ is very small (in the range of 15-30meV) and it can be considered as metallic materials at room temperature. However, for AGNR$_M$, this family exhibits a relatively large $E_g$ and is considered as semiconductor materials, similar to the other two families. According to Ref. 8, $E_g$ of this family is strongly affected by the edge effect of AGNR$_M$. However for AGNR$_B$, there exists an electron-electron interaction between the layers at the edges and hence the edge effect diminishes as compared to AGNR$_M$, which leads to the metallic nature for this family of AGNR$_B$. For $N$=3p+1, the $E_g$ of AGNR$_B$ is larger than the $N$=3p family at small $W$. At larger $W$, the $E_g$ of

$N=3p+1$ decreases dramatically and becomes similar to that of $N=3p$. It indicates that the edge exert less effect on $E_g$ when the $AGNR_B$ width is over 2.4nm. In addition to the variation in $W$, the $E_g$ of $AGNR_B$ is also affected by the interlayer distance ($D$) and the $E_g$ dependency on $D$ for $N=5$, 6 and 7 is presented in Fig. 3(b). For $D$ larger than the respective $D_{op}$, $E_g$ increases as $D$ increases, with $N=3p+1$ showing the largest increment. Conversely, as $D$ decreases, $E_g$ decreases. Therefore, it indicated that the electronic properties of $AGNR_B$ can be controlled by the width of the ribbon as well as the interlayer distance. This opens up another possible avenue for device design using bandgap engineering.

Finally, we explore the edge doping effects of $AGNR_M$ and $AGNR_B$ on their electronic structures. Edge doping concentration is defined as the number of dopant atoms, boron (triangle) and nitrogen (diamond), per unit length per layer of $AGNR_M$[13]. The most energetically favorable doping site is along the edges of the ribbons[12] and it is shown in the boxed regions of Fig. 1(a). In order to understand the change in Fermi level, $N=7$ is chosen for both $AGNR_M$ (red) and $AGNR_B$ (blue). As shown in Fig. 4, the Fermi level of both $AGNR_M$ and $AGNR_B$ decreases as the concentration of boron increases, which corresponds to the formation of a *p*-type semiconductor. Similarly, as the concentration of nitrogen increases, the Fermi level of both AGNR systems increases, showing an *n*-type semiconductor. This confirms that both $AGNR_M$ and $AGNR_B$ can form doped semiconductor with suitable atoms attached at the edges. Furthermore, the $E_g$ values for the doped AGNR systems are calculated and it is found that the $E_g$ of $AGNR_M$ decreases by 11-17% while that of $AGNR_B$ decreases by 4-10%. Therefore, $AGNR_B$ exhibits more stable electronic property than $AGNR_M$.

In this Letter, we explore the fundamental of energy bandgap of armchair-edge bilayer GNR ($AGNR_B$) based on the first-principles calculations. Our simulations show that the optimum interlayer distance ($D$) is about 3.1Å for 7-AGNR$_B$, which is different from the interlayer distance ($D{\approx}3.3$Å) of bilayer graphene due to the difference in charge distribution between the layers. We also examine the width dependency of the energy gap ($E_g$) of AGNR$_B$ and compared with that of monolayer GNR (AGNR$_M$). The $E_g$ of AGNR$_B$ is generally smaller than AGNR$_M$ and unlike AGNRM, one group ($N$=3p+2) of the $E_g$ trends shows metallic behavior. Furthermore, we investigated the effect of interlayer distance on $E_g$ of AGNR$_B$ and found that $E_g$ strongly depends on $D$. It indicates possible application in devices where $E_g$ can be tuned by varying $D$. The edge doping concentration in AGNR$_M$ and AGNR$_B$ is also investigated and it shows that the Fermi level is changed by the type and concentration of the dopant, forming $p$-type (with boron) and $n$-type (with nitrogen) semiconductor. The study on the electronic structure of AGNR$_B$ is an important precursor to the study of future AGNR$_B$ devices. Although the $E_g$ of AGNR$_B$ is smaller than that of AGNR$_M$, recent study[18] has shown that an externally applied electric field can open up the $E_g$ of a bilayer graphene and this may also be applicable to AGNR$_B$.

This work was supported by the Ministry of Education of Singapore under Grant Nos. R-263-000-416-112 and R-263-000-416-133, and National Research Foundation CRP grant.

Figure Captions

Fig. 1 (a) The honeycomb structure of a 7-AGNR$_M$. (b)The stacking configuration of a graphene bilayer. The top carbon layer (gold) is arranged such that one atom is positioned in the centre of the hexagon of the bottom layer (blue). (c)The atomic structure of a 7-AGNR$_B$ with $D$ representing the interlayer distance.

Fig. 2 (a) Total energy of a bilayer graphene and 7-AGNR$_B$ as a function of the interlayer distance ($D$). Total energy is normalized as a ratio to the lowest value of the respective structure in the data. The optimum $D$ (at lowest energy) of a bilayer graphene and 7-AGNR$_B$ are 3.3Å and 3.1Å, respectively. Moreover, as $D>4.9$Å, total energy in both cases shows saturation and it indicates that bilayer graphene sheets and GNR$_B$ may behave like the monlayer grahene sheets and GNR$_M$. Insert shows that the optimum $D$ increases as width of the AGNR$_B$ increases. (b) and (c) show the charge density of the planes in the middle of the bilayer graphene and 7-AGNR$_B$, respectively. The unit of the color gradient bar is 1/Bohr$^3$. The difference in charge distribution in these two cases causes the deviation in optimum $D$ in Fig. 2(a).

Fig. 3 (a) Energy gap ($E_g$) as a function of the width of AGNR$_B$ for interlayer distance ($D$)=6.5Å (hollow points) and the respective optimum $D$ (solid points). Diamond, circle, and square points represent the three different families of $N$=3p, 3p+1 and 3p+2, respectively. Unlike AGNR$_M$, the family of $N$=3p+2 in AGNR$_B$ shows almost zero bandgap for any width. The dot-dash lines show the three trends in AGNR$_M$ which coincide with the $E_g$ trends of AGNR$_B$ when $D$=6.5Å. It demonstrates that AGNR$_B$ with $D$=6.5Å shows the same behaviors as AGNR$_M$. (b) Dependence of $E_g$ on $D$ of the AGNR$_B$ for $N$=5, 6 and 7. Electronic structure of AGNR$_B$ strongly depends on its interlayer distance.

Fig. 4 Fermi level *vis-a-vis* different boron (triangle) and nitrogen (diamond) doping concentrations for 7-AGNR$_M$ (blue) and 7-AGNR$_B$ (red). The dot-dash and dotted lines are the original Fermi level of undoped 7-AGNR$_M$ and 7-AGNR$_B$ respectively. It can be clearly seen that *p*-type (with boron) and *n*-type (nitrogen) semiconductors are formed with increasing dopant concentration. As doping concentration increases, bandgap of all four cases also decreases.

Figure 1, Lam, K.T. and Liang, G.C.

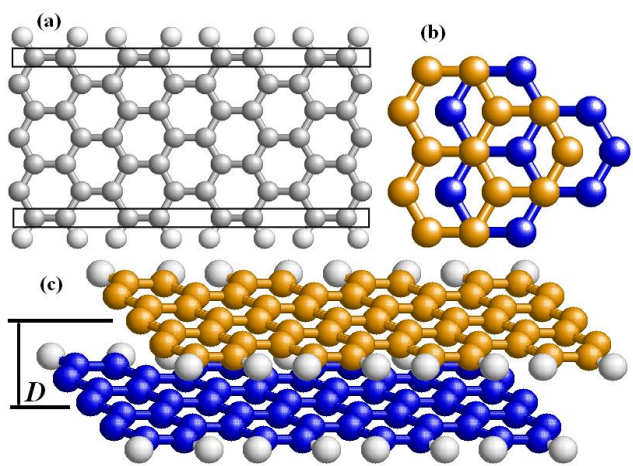

Figure 2, Lam, K.T. and Liang, G.C.

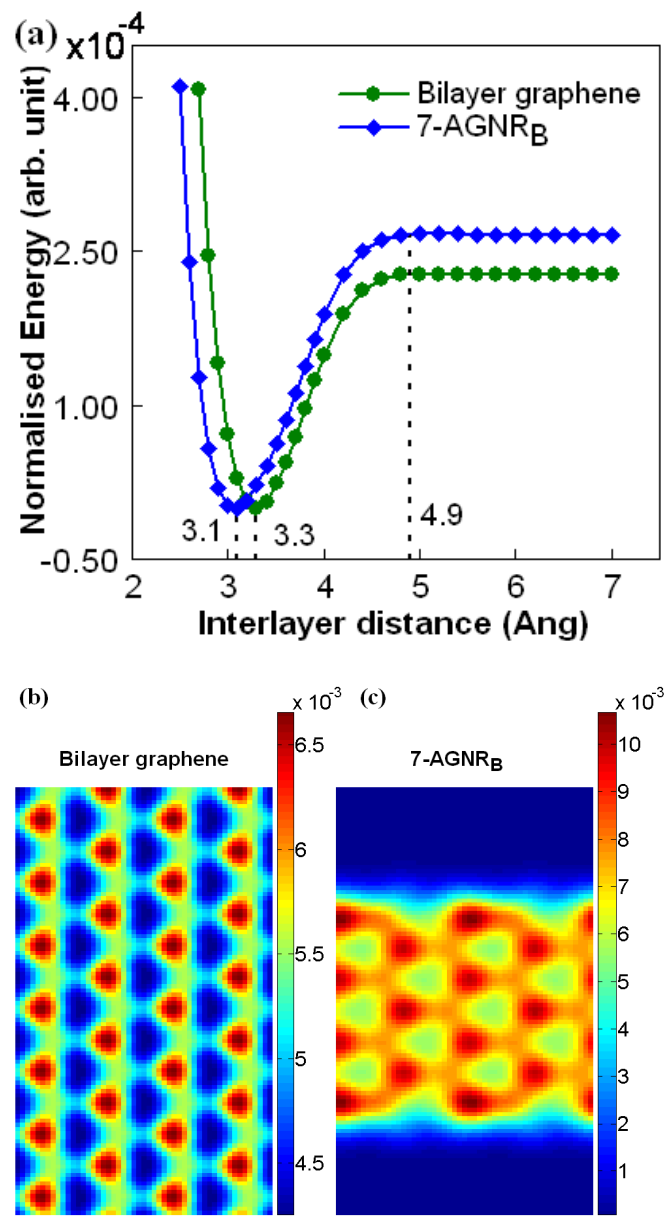

Figure 3, Lam, K.T. and Liang, G.C.

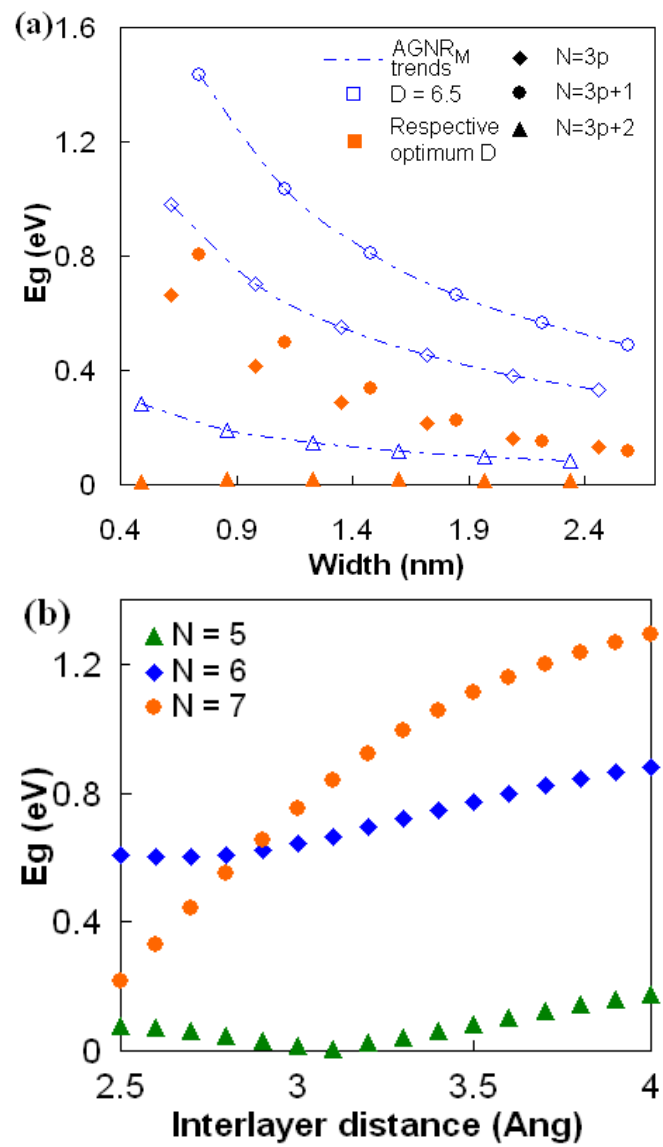

Figure 4, Lam, K.T. and Liang, G.C.

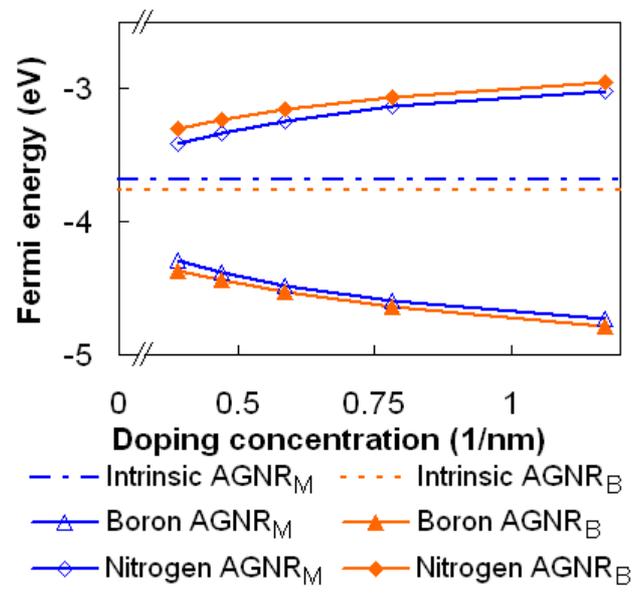